\documentstyle[amssymb,preprint,aps]{revtex}

\tightenlines

\begin{document}
\draft
\title{Adsorption of a random heteropolymer with self-interactions onto an
interface }
\author{Valerii A. Brazhnyi$^{1,2}$ and Semjon Stepanow$^{1}$}
\address{$^{1}$Martin-Luther-Universit\"{a}t Halle-Wittenberg, Fachbereich\\
Physik, D-06099 Halle, Germany}
\address{$^{2}$Sumy State University, 2, Rimskii-Korsakov Str., 40007 Sumy,\\
Ukraine}
\date{\today }
\maketitle

\begin{abstract}
We consider the adsorption of a random heteropolymer onto an interface
within the model by Garel et al. \cite{gareletal89} by taking into account
self-interactions between the monomers. Within the replica trick and by
using a self-consistent preaveraging procedure we map the adsorption problem
onto the problem of binding state of a quantum mechanical Hamiltonian. The
analysis of the latter is treated within the variational method based on the
2-nd Legendre transform. We have found that self-interactions favor the
localization. The effect is intensified with decrease of the temperature.
Within a model without taking into account the repulsive ternary
monomer-monomer interactions we predict a reentrant localization transition
for large values of the asymmetry of the heteropolymer and at low enough
temperatures.
\end{abstract}


\section*{Introduction}

The behavior of heteropolymers at interface between two immiscible
(incompatible) solvents has been intensively studied in recent years \cite
{gareletal89}-\cite{sommer/daoud95} since it has an obvious importance in
biological applications (proteins and membranes) \cite{garel/orland/pitard}
and application in different fields of industry such as biosensor, pattern
recognition applications, glues, paints etc. \cite{Napper}. Experiments \cite
{daietal/prl94}, \cite{brownetal93} and numerical simulations \cite
{sommer/peng/blumen}, \cite{chen/jcp99} have shown that a random
heteropolymer may localize, reinforcing the interface between two
incompatible solvents and reduce interfacial tension. Recent theoretical
efforts have been devoted to understand the fundamental physical mechanism
governing the localization of a random copolymer onto an interface \cite
{stepanowetal98}-\cite{denesyuk01}.

In the simple model introduced by Garel et al. \cite{gareletal89} only the
interaction of the monomers with the solvent, but not the self-interactions
between the monomers were taken into account. Considering a $A-B$ copolymer
at the $A-B$ interface $A$ monomers prefer to be in the $A$-half-plane while 
$B$ monomers prefer to be in $B$-half-plane. Obviously there is a sort of
frustration in such system because the complete separation of $A$ monomers
into one-half and $B$ monomers into another half plane is forbidden by
polymer bonds. The analysis performed in \cite{gareletal89} showed that the
localized state of a random heteropolymer chain in the presence of a
selective interface can be imagined as consisting of blobs with majority of $%
A$ monomers and minority of $B$ monomers in the $A$-half-plane and vice
versa for the blobs in the $B$-half-plane. It was shown that random
copolymer always localizes for statistically symmetric heteropolymer,
whereas a delocalization transition was found if the heteropolymer is
asymmetric. The heterogeneity in the chemical structure of the polymer,
which results in self-interactions between the monomers, may have a
considerable impact on their bulk thermodynamic behavior \cite
{fredrickson/milner91}, which consists in a segregation into $A$-rich and $B$%
-rich domains. In the case of a single heteropolymer in solvent the
self-interactions favor the collapse of the heteropolymer \cite
{garel/leibler/orland94}. The self-interactions between monomers may play an
important role for adsorption as it was noticed in \cite
{milner/fredrickson95}, \cite{gutman/chakraborty1/4}. However, the influence
of self-interactions on the localization behavior of the random
heteropolymer is poorly understood. The effect of the excluded volume in the
adsorption of a heteropolymer was recently investigated in \cite
{dwyeretal/jcp00}. In the present paper we will consider the influence of
the direct monomer-monomer interactions on the localization behavior of a
random heteropolymer onto\ an interface.

The article is organized as follows. In Section II we will introduce to the
model and the formalism we use. Section III contains our results and Section
IV conclusions.

\section{Model and Formalism}

We study the problem of a random heteropolymer by using the model proposed
in \cite{gareletal89}. Let us consider a long two-letter $A-B$ heteropolymer
chain in the presence of the $A^{\prime }-B^{\prime }$ interface between two
incompatible solvents. Alternation of different type of monomers along the
chain is assumed to be random, and each monomer is assumed to interact with
an external potential, which takes positive and negative values depending on
the position with respect to the interface. The partition function of the
polymer in the presence of the interface is given by the following Edwards
functional\ integral \cite{edwards65} over the trajectories ${\bf r}(s)$ ($%
0\leq s\leq N$) of the polymer chain 
\begin{eqnarray}
Z\{\zeta (s)\} &=&\int D{\bf r}(s)\exp \left\{ -\frac{1}{2a^{2}}%
\displaystyle\int %
\nolimits_{0}^{N}ds\left( \frac{\partial {\bf r}}{\partial s}\right) ^{2}+w%
\displaystyle\int %
\nolimits_{0}^{N}ds\;\zeta (s){\rm sgn}[z(s)]+\right.  \nonumber \\
&&\left. \chi _{0}%
\displaystyle\int %
\nolimits_{0}^{N}ds%
\displaystyle\int %
\nolimits_{0}^{N}ds^{\prime }\;\zeta (s)\zeta (s^{\prime })\delta \left( 
{\bf r}(s)-{\bf r}(s^{\prime })\right) \right\} ,  \label{eqn1}
\end{eqnarray}
where $a$ is the Kuhnian segment length, $z(s)$ is the Cartesian component
of ${\bf r}(s)$ transversal to the interface, $w$ and $\chi _{0}$\ are
measured in units of $k_{B}T$. The first term in the exponential of (\ref
{eqn1}) corresponds to the elastic energy of the polymer chain, the second
one describes the monomer interaction with the medium, which is governed by
the random parameter $\zeta (s)$, which is assumed to be Gaussian
distributed with the distribution function

\begin{equation}
P\{\zeta (s)\}\varpropto \exp \left[ -\frac{1}{2\Delta _{0}}%
\int_{0}^{N}\left( \zeta (s)-\zeta _{0}\right) ^{2}ds\right] .  \label{eqn2}
\end{equation}
The distribution function (\ref{eqn2}) of the random ``charges'' $\zeta (s)$
is completely characterized by its two moments, $\left\langle \zeta
(s)\right\rangle =\zeta _{0}$ with $\zeta _{0}$ related to the asymmetry in
the composition of the copolymer, and $\left\langle \zeta (s)\zeta
(s^{\prime })\right\rangle =\zeta _{0}^{2}+\Delta _{0}\delta (s-s^{\prime })$
where $\Delta _{0}$ being the variance of the distribution. The interaction
potential is chosen as a step function. Such a choice is legitimate, if the
interface width is much smaller than the size of an average excursion loop.
The last term in Eq.(\ref{eqn1}) describes the short-rang interaction
between monomers, where $\chi _{0}$ is the effective interaction potential
(the second virial coefficient). The sign of the last term in (\ref{eqn1})
is chosen so that for $\chi _{0}>0$ the like monomers attract each other
while the unlike monomers repel each other. The model described by Eq.(\ref
{eqn1}) admit that the self-interactions occur independent of whether the
monomers are on the left or on the right side of the interface. This may be
the case if even the favorable solvent for a given type of monomer is
slightly poor. If the favorable solvent is comprised of the same monomers,
then the self-interactions between the like monomers in their own medium are
expected to be zero. The generalization of the model where the
self-interactions are switched off, if they are in their own medium is
possible. The effect of ternary interactions, which will prevent the
collapse of the blobs of the adsorbed heteropolymer due to attractive
self-interactions, will be considered in a separate paper.

The random ``charges'' $\zeta (s)$ in (\ref{eqn1}) are considered as
quenched variables, so that in order to obtain the free energy one has to
average $\ln (Z)$ over all possible realizations of monomer sequences. For
this purpose we use the replica trick\ consisting in introduction of $n$
copies of the system with the same quenched variables $\zeta (s)$, and using
the identity $\ln (Z)=\lim_{n\rightarrow 0}(Z^{n}-1)/n$ in averaging over $%
\zeta (s)$. Thus, at the intermediate stage we consider the average $%
\left\langle Z^{n}\right\rangle $ where $\left\langle ...\right\rangle $
means average with the distribution function (\ref{eqn2}). The partition
function $Z^{n}$ can be written as 
\begin{equation}
Z^{n}=\int \prod\limits_{\alpha =1}^{n}D{\bf r}_{\alpha }(s)\exp \left\{
-H_{0}-H_{int}\right\} ,  \label{eqn4}
\end{equation}
where 
\begin{equation}
H_{0}=\frac{1}{2a^{2}}%
\displaystyle\int %
\nolimits_{0}^{N}ds\sum\limits_{\alpha =1}^{n}\left( \frac{\partial {\bf r}%
_{\alpha }}{\partial s}\right) ^{2},  \label{eqn5}
\end{equation}

\begin{equation}
H_{int}=-w%
\displaystyle\int %
\nolimits_{0}^{N}ds\;\zeta (s)\sum\limits_{\alpha =1}^{n}{\rm sgn}[{\bf r}%
_{\alpha }(s)]-\chi _{0}%
\displaystyle\int %
\nolimits_{0}^{N}ds%
\displaystyle\int %
\nolimits_{0}^{N}ds^{\prime }\;\zeta (s)\zeta (s^{\prime
})\sum\limits_{\alpha =1}^{n}\delta \left( {\bf r}_{\alpha }(s)-{\bf r}%
_{\alpha }(s^{\prime })\right) .  \label{eqn6}
\end{equation}
To average over $\zeta (s)$ in (\ref{eqn4}) we expand (\ref{eqn4}) in Taylor
series in powers of $H_{int}$, carry out the average, and reexponentiate the
obtained expression. This results in writing the result of the average in
the exponential as

\begin{equation}
\ln \left( \left\langle \exp \left\{ -H_{int}\right\} \right\rangle \right)
=-\left\langle H_{int}\right\rangle +\frac{1}{2}\left( \left\langle
H_{int}^{2}\right\rangle -\left\langle H_{int}\right\rangle ^{2}+...\right) .
\label{eqn7}
\end{equation}
Restricting the expansion in (\ref{eqn7}) to quadratic terms in $H_{int}$ we
obtain 
\begin{equation}
\left\langle Z^{n}\right\rangle =\int \prod\limits_{\alpha =1}^{n}D{\bf r}%
_{\alpha }(s)\exp \left\{ -%
\displaystyle\int %
\nolimits_{0}^{N}{\cal L}_{n}\,\,ds\right\} ,  \label{eqn7a}
\end{equation}
with

\begin{eqnarray}
{\cal L}_{n} &=&\frac{1}{2a^{2}}\sum\limits_{\alpha =1}^{n}(\frac{\partial 
{\bf r}_{\alpha }(s)}{\partial s})^{2}-w\zeta _{0}\sum\limits_{\alpha =1}^{n}%
{\rm sgn}({\bf r}_{\alpha }(s))-\chi _{0}\zeta _{0}^{2}\sum\limits_{\alpha
=1}^{n}%
\displaystyle\int %
\nolimits_{0}^{N}\delta \left( {\bf r}_{\alpha }(s)-{\bf r}_{\alpha
}(s^{\prime })\right) \,ds^{\prime }-  \nonumber \\
&&\frac{1}{2}w^{2}\Delta _{0}\sum\limits_{\alpha ,\beta =1}^{n}{\rm sgn}(%
{\bf r}_{\alpha }(s)){\rm sgn}({\bf r}_{\beta }(s))-2w\chi _{0}\zeta
_{0}\Delta _{0}\sum\limits_{\alpha ,\beta =1}^{n}{\rm sgn}({\bf r}_{\alpha
}(s))%
\displaystyle\int %
\nolimits_{0}^{N}\delta \left( {\bf r}_{\beta }(s)-{\bf r}_{\beta
}(s^{\prime })\right) \,ds^{\prime }-  \nonumber \\
&&2\chi _{0}^{2}\zeta _{0}^{2}\Delta _{0}\sum\limits_{\alpha ,\beta =1}^{n}%
\displaystyle\int %
\nolimits_{0}^{N}ds^{\prime }\,%
\displaystyle\int %
\nolimits_{0}^{N}ds^{\prime \prime }\delta \left( {\bf r}_{\alpha }(s)-{\bf r%
}_{\alpha }(s^{\prime })\right) \delta \left( {\bf r}_{\beta }(s)-{\bf r}%
_{\beta }(s^{\prime \prime })\right) \,-  \nonumber \\
&&\chi _{0}^{2}\Delta _{0}^{2}\sum\limits_{\alpha ,\beta =1}^{n}%
\displaystyle\int %
\nolimits_{0}^{N}\delta \left( {\bf r}_{\alpha }(s)-{\bf r}_{\alpha
}(s^{\prime })\right) \,\delta \left( {\bf r}_{\beta }(s)-{\bf r}_{\beta
}(s^{\prime })\right) \,ds^{\prime }.  \label{eqn8}
\end{eqnarray}
Notice that ${\cal L}_{n}$ contains more than one integration over the
contour length. Due to this it is not possible to reduce ${\cal L}_{n}$ to a
quantum mechanical Hamiltonian as it is the case, if only a single
integration over the contour length appears (see for example \cite
{stepanowetal98}). In the following we will preaverage (\ref{eqn8}) in order
to reduce it to a quantum mechanical Hamiltonian. For this purpose let us
consider the quantities

\begin{eqnarray}
\rho ({\bf r}_{\alpha }) &=&%
\displaystyle\int %
\nolimits_{0}^{N}\delta \left( {\bf r}_{\alpha }-{\bf r}_{\alpha }(s^{\prime
})\right) \,ds^{\prime },  \label{eqn9} \\
q({\bf r}_{\alpha },{\bf r}_{\beta }) &=&%
\displaystyle\int %
\nolimits_{0}^{N}\delta \left( {\bf r}_{\alpha }-{\bf r}_{\alpha }(s^{\prime
})\right) \,\delta \left( {\bf r}_{\beta }-{\bf r}_{\beta }(s^{\prime
})\right) \,ds^{\prime },  \label{eqn10}
\end{eqnarray}
where $\rho ({\bf r}_{\alpha })$ is the microscopic density associated with
the polymer and $q({\bf r}_{\alpha },{\bf r}_{\beta })$ is the replica
overlap parameter \cite{garel/leibler/orland94}. We will preaverage $\rho (%
{\bf r}_{\alpha })$ and $q({\bf r}_{\alpha },{\bf r}_{\beta })$ in Eq.(\ref
{eqn8}) over ${\bf r}_{\alpha }(s^{\prime })$ and ${\bf r}_{\beta
}(s^{\prime })$ according to

\begin{eqnarray}
\left\langle \rho ({\bf r}_{\alpha })\right\rangle &=&%
\displaystyle\int %
\nolimits_{0}^{N}ds\int dZ\int dZ^{\prime }\,G(Z,N;z,s)G(z,s;Z^{\prime
},0)\times  \nonumber \\
&&\int dR_{\shortparallel }\int dR_{\shortparallel }^{\prime
}G_{0}(R_{\shortparallel },N;r_{\shortparallel },s)G_{0}(r_{\shortparallel
},s;R_{\shortparallel }^{\prime },0)  \label{eqn13}
\end{eqnarray}
and similar for $q({\bf r}_{\alpha },{\bf r}_{\beta })$. $%
G_{0}(R_{\shortparallel },N;r_{\shortparallel },s)$ in (\ref{eqn13}) is the
free Green's function of the longitudinal degrees of freedom of the polymer.
Notice that the average over the longitudinal coordinates of the polymer
with the unperturbed Green's function is an approximation. We use the
spectral representation of the transversal Green's function in Eq.(\ref
{eqn13}) $G(z,N;z^{\prime },0)$ through the eigenfunctions $\psi _{k}(z)$

\begin{equation}
G(z,z^{\prime };N)=\sum\limits_{k}e^{-N\varepsilon _{k}}\psi _{k}(z)\psi
_{k}^{\ast }(z^{\prime })\,.  \label{eqn14}
\end{equation}
The Green's function $G(z,z^{\prime };N)$ satisfies for $N>0$ the
differential equation

\begin{equation}
-\frac{\partial G}{\partial N}=-\frac{a^{2}}{6}\partial _{z}^{2}G+\frac{U(z)%
}{T}G,  \label{eqn15}
\end{equation}
which is remarkably similar to the Schr\"{o}dinger equation for a quantum
particle in an external potential \cite{edwards65}

\[
i\hbar \frac{\partial \psi }{\partial t}=-\frac{\hbar ^{2}}{2m}\nabla
^{2}\psi +U(r)\,\psi . 
\]
The mapping of the latter onto Eq.(\ref{eqn15}) occurs by using the
following replacements: $t\rightarrow iN$, $\hbar \rightarrow T$, $%
a^{2}/T\rightarrow 3/m$.

In the case of a discrete spectrum with the energy of the ground state being
negative, the main contribution in the series (\ref{eqn14}) for large $N$
originates from the ground state (ground state dominance) \cite
{degennes/book}. The preaveraging of $\rho ({\bf r}_{\alpha })$ and $q({\bf r%
}_{\alpha },{\bf r}_{\beta })$ given by Eqs.(\ref{eqn9}-\ref{eqn10})
according to (\ref{eqn13}) in the approximation of the ground state
dominance gives

\begin{eqnarray}
\left\langle \rho ({\bf r}_{\alpha })\right\rangle &\approx &\sigma \,\left|
\psi _{k_{0}}(z_{\alpha })\right| ^{2},  \label{eqn16} \\
\left\langle q({\bf r}_{\alpha },{\bf r}_{\beta })\right\rangle &\approx
&\sigma \,\left| \psi _{k_{0}}(z_{\alpha })\right| ^{2}\left| \psi
_{k_{0}}(z_{\beta })\right| ^{2},  \label{eqn17}
\end{eqnarray}
where $\sigma =N/S$ is an average monomer density per surface area \cite
{milner/witten88}. We expect that the ratio $\sigma $ is finite for large $N$
and $S$.

The substitution of (\ref{eqn16}), (\ref{eqn17}) into the Hamiltonian (\ref
{eqn8})\ gives the effective replica Hamiltonian as 
\begin{eqnarray}
H_{n} &=&-D\sum\limits_{\alpha =1}^{n}\nabla _{z}^{2}-w\zeta
_{0}\sum\limits_{\alpha =1}^{n}{\rm sgn}(z_{\alpha })-\chi _{0}\zeta
_{0}^{2}\sigma \sum\limits_{\alpha =1}^{n}\,\left| \psi (z_{\alpha })\right|
^{2}-  \nonumber \\
&&\frac{1}{2}w^{2}\Delta _{0}\sum\limits_{\alpha ,\beta =1}^{n}{\rm sgn}%
(z_{\alpha }){\rm sgn}(z_{\beta })-2w\chi _{0}\zeta _{0}\Delta _{0}\sigma
\sum\limits_{\alpha ,\beta =1}^{n}{\rm sgn}(z_{\alpha })\,\left| \psi
(z_{\beta })\right| ^{2}-  \nonumber \\
&&(2\zeta _{0}^{2}\sigma +\Delta _{0})\chi _{0}^{2}\Delta _{0}\sigma
\sum\limits_{\alpha ,\beta =1}^{n}\,\left| \psi (z_{\alpha })\right|
^{2}\left| \psi (z_{\beta })\right| ^{2},  \label{eqn18}
\end{eqnarray}
where we have introduced the notation $D=a^{2}/2$. Due to the average over
the longitudinal degrees of freedom of the polymer by using the unperturbed
Green's function, the problem becomes one dimensional.

The investigation of the adsorption of a random heteropolymer chain is
equivalent to the study of the ground state of the Hamiltonian $H_{n}$ given
by Eq.(\ref{eqn18}). Without taking into account the self-interactions, the
case which is obtained from Eq.(\ref{eqn18}) by putting $\chi _{0}=0$,
Stepanow et al. \cite{stepanowetal98} applied a novel variational principle
for Green's function associated with the Hamiltonian $H_{n}$. The latter
generalizes the well-known Rayleigh-Ritz method in Quantum Mechanics for
nonstationary states. The variational principle for the Green's function can
be outlined as follows \cite{stepanowetal98}. The start point is the Dyson
equation for the Green's function $G$

\begin{equation}
0=-G^{-1}+G_{0}^{-1}+H_{n}^{int},  \label{eqn19}
\end{equation}
which is considered as a stationarity condition $\delta F(G)/\delta G=0$ of
a functional $F(G)$, which is found in a straightforward way as

\begin{equation}
F(G)=-{\rm tr}\,\ln (G)+{\rm tr}\,G_{0}^{-1}G+{\rm tr}\,H_{n}^{int}G,
\label{eqn20}
\end{equation}
where the bare Green's function is defined as $G_{0}^{-1}=\omega +H_{0}$,
with $H_{0}$ being the unperturbed part of the Hamiltonian, $H_{n}^{int}$ is
the interaction part of the Hamiltonian, and $\omega $ is Laplace conjugate
to the chain's length $N$. The functional $F(G)$\ given by Eq.(\ref{eqn20})
is the particular case of the generating functional of the 2-nd Legendre
transform in Quantum Field Theory and Statistical Physics \cite{dedominicis}%
- \cite{stepanowetal96}{\it .} Notice that without preaveraging of $\rho (%
{\bf r}_{\alpha })$ and $q({\bf r}_{\alpha },{\bf r}_{\beta })$ given by
Eqs.(\ref{eqn9}-\ref{eqn10}) the problem under consideration does not reduce
to a quantum mechanical problem and instead of (\ref{eqn20}) we have to use
the extremum principle associated with the second Legendre transform \cite
{DobrEr}-\cite{stepanowetal96}. Assuming the ground state dominance we
choose the $n$-replica trial Green's function as 
\begin{equation}
G(k_{1,}k_{2};z,z^{\prime };t)=%
\mathop{\displaystyle\prod}%
\limits_{\alpha =1}^{n}\exp (-\varepsilon _{k}t)\psi (k_{1,}k_{2};z_{\alpha
})\psi (k_{1,}k_{2};z_{\alpha }^{\prime }),  \label{eqn21}
\end{equation}
where the trial wave function is chosen as

\begin{equation}
\psi (k_{1,}k_{2};z_{\alpha })=\sqrt{\frac{2k_{1}k_{2}}{k_{1}+k_{2}}}\left(
e^{-k_{1}z_{\alpha }}\vartheta (z_{\alpha })+e^{k_{2}z_{\alpha }}\vartheta
(-z_{\alpha })\right) .  \label{eqn22}
\end{equation}
Notice that the energy $\varepsilon _{k}=-Dk_{2}^{2}$ is negative, and is a
function of $k_{2}$. Computing the terms in (\ref{eqn20}) by using (\ref
{eqn21}), (\ref{eqn22}) gives the extremum functional as

\begin{eqnarray}
F(k_{1,}k_{2}) &=&\ln (\omega +n\varepsilon _{k})+\frac{n(Dk_{1}k_{2}-%
\varepsilon _{k})}{\omega +n\varepsilon _{k}}-\frac{nw\zeta _{0}}{\omega
+n\varepsilon _{k}}\left( \frac{k_{2}-k_{1}}{k_{1}+k_{2}}\right)  \nonumber
\\
&&-\frac{n}{2}\frac{\Delta _{0}w^{2}}{\omega +n\varepsilon _{k}}-\frac{n(n-1)%
}{2}\frac{\Delta _{0}w^{2}}{\omega +n\varepsilon _{k}}\left( \frac{%
k_{2}-k_{1}}{k_{1}+k_{2}}\right) ^{2}  \nonumber \\
&&-\frac{n\chi _{0}\zeta _{0}^{2}\sigma }{\omega +n\varepsilon _{k}}\frac{%
k_{1}k_{2}}{k_{1}+k_{2}}-\frac{n^{2}w\chi _{0}\zeta _{0}\Delta _{0}\sigma }{%
\omega +n\varepsilon _{k}}\frac{2k_{1}k_{2}}{k_{1}+k_{2}}\left( \frac{%
k_{2}-k_{1}}{k_{1}+k_{2}}\right)  \nonumber \\
&&-\frac{(n+3n^{2})}{12}\frac{(2\zeta _{0}^{2}\sigma +\Delta _{0})\chi
_{0}^{2}\Delta _{0}\sigma }{\omega +n\varepsilon _{k}}\left( \frac{%
2k_{1}k_{2}}{k_{1}+k_{2}}\right) ^{2}.  \label{eqn23}
\end{eqnarray}

The stationarity conditions

\begin{equation}
\partial F/\partial k_{1}=\partial F/\partial k_{2}=0,  \label{eqn24}
\end{equation}
give in the limit $n=0$ the following equations 
\begin{eqnarray}
&&3k_{1}^{3}+9k_{1}^{2}k_{2}+9k_{1}k_{2}^{2}+3k_{2}^{3}+6\zeta
(k_{1}+k_{2})+6\Delta (k_{1}-k_{2})-  \nonumber \\
&&3\chi \zeta ^{2}\sigma k_{2}(k_{1}+k_{2})-2(2\zeta ^{2}\sigma +\Delta
)\chi ^{2}\Delta \sigma k_{1}k_{2}^{2}=0,  \label{eqn25}
\end{eqnarray}

\begin{eqnarray}
&&3k_{1}^{3}k_{2}+9k_{1}^{2}k_{2}^{2}+9k_{1}k_{2}^{3}+3k_{2}^{4}+3\zeta
(k_{1}^{2}-k_{2}^{2})-6\Delta k_{1}k_{2}-  \nonumber \\
&&3\chi \zeta ^{2}\sigma k_{1}k_{2}(k_{1}+k_{2})-(2\zeta ^{2}\sigma +\Delta
)\chi ^{2}\Delta \sigma k_{1}k_{2}^{2}=0,  \label{eqn26}
\end{eqnarray}
where we have introduced new quantities $\Delta =\Delta _{0}w^{2}$, $\zeta
=\zeta _{0}w$, $\ \chi =\chi _{0}/w^{2}$ and have put $D=1$. Solution of
this system of equations gives us the localization lengths $\xi _{1}=1/k_{1}$
\ and $\xi _{2}=1/k_{2}$, which describe the localization of the random
heteropolymer onto the interface.

\section{Results}

The present problem was studied in \cite{stepanowetal98} (see also \cite
{ganesan/brenner99}-\cite{denesyuk01}) without taking into account the
self-interactions between the monomers. It was found in \cite{stepanowetal98}
that the localization-delocalization transition occurs at the temperature $%
T_{c}=\frac{2\Delta _{0}}{3\zeta _{0}}$, where the parameter $k_{1}$ becomes
zero and localization length $\xi _{1}$ becomes infinite. The value of $%
T_{c} $ coincides exactly with that found in \cite{monthus/epb00} by using a
different method. The condition $k_{1}=0$ means that the heteropolymer
delocalizes in the right hand half plane ($z>0$).

Unfortunately, the equations (\ref{eqn25}), (\ref{eqn26}) cannot be solved
analytically, so that we analyze them numerically. First we have examined
the influence of the monomer-monomer interaction on the localization
lengths. In Fig.1,2 we present the dependence of \ $k_{1}$ and $k_{2}$ as a
functions of the asymmetry parameter $\zeta _{0}$ for different values of
the parameter $\chi _{0}$. Fig.1 corresponds to $\chi _{0}=0$ and recovers
the results of \cite{stepanowetal98} with values $k_{1}^{0}=\sqrt{6\Delta }%
/9 $ and $k_{2}^{0}=2k_{1}^{0}$ for statistically symmetric heteropolymer, $%
\zeta _{0}=0$. At the critical value of the asymmetry $\zeta
_{0}^{c}=2\Delta /3$, the parameter $k_{1}$ becomes zero, so that the
heteropolymer delocalizes in the right hand half-plane. The increase of the
parameter $\chi _{0}$ (Fig.2) leads to the increase of the critical value of
the asymmetry parameter $\zeta _{0}^{1}$ at which the delocalization
transition takes place. This means that the self-interactions favor the
localization of the heteropolymer. This can be explained qualitatively as
follows. The self-interactions being attractive results in a decrease of the
size of the blobs on both sides of the interface, which result in a decrease
of their repulsion from the interface. The self-interactions influence to
great extent the larger blobs, i.e. the blobs which are on the right hand
side of the interface, while as in the case of the excluded volume
interactions the effect of interactions is proportional to $\chi \sqrt{N}$
with $N$ being the number of monomers in the blob.\ At given $\chi _{0}>0$
and large enough $\zeta _{0}$ and $T$ (see Fig.2, curve 2) the parameter $%
k_{1}$ becomes again positive that means that in this range of parameters
the polymer is localized. Thus, as a function of the asymmetry parameter $%
\zeta _{0}^{1}<\zeta _{0}<\zeta _{0}^{2}$ with $\zeta _{0}^{1}$ and $\zeta
_{0}^{2}$ depending on $\chi $, there is window where the polymer is
delocalized. For $\zeta _{0}<\zeta _{0}^{1}$ the self-interactions favor the
localization driving the critical value $\zeta _{0}^{c}$ to larger values.
It is surprising that for $\zeta _{0}>\zeta _{0}^{2}$ the polymer localizes
again. We interpret this as follows. We expect that in the model under
consideration it is reasonable to impose a restriction on the maximal value
of the asymmetry parameter $\zeta _{0}^{\max }$, which is of order of $%
\Delta _{0}$ \cite{gutman/chakraborty1/4}, so that $\zeta _{0}<$ $\zeta
_{0}^{\max }$. Although at $\zeta _{0}=\zeta _{0}^{\max }$ the heteropolymer
is on average homopolymer, due to the difference of the variance $\Delta
_{0} $ from zero the typical polymer is still heterogeneous. The
self-interactions between the monomers in Hamiltonian (\ref{eqn18}), which
are attractive between the monomers of the same kind and repulsive between
the monomers of different kind, have the tendency to cause a microphase
separation between the $A$ and $B$ monomers at large values of asymmetry
parameter $\zeta _{0}$. Then, the polymer can be imagined as consecutive
units of microphase separated blobs, each of them prefers to be on the left
or on the right with respect to the interface. This is expected to favor the
localization of the heteropolymer. We expect that taking into account the
incompressibility of the blobs will disfavor the reentrant transition.
However, the prediction that the self-organization of the random
heteropolymer due to self-interactions will favor the localization will
remain valid.

Notice that both $k_{1}$ and $k_{2}$ increase with increase of $\zeta _{0}$
and become equal at some $\zeta _{0}$ (curve 1 in Fig.2). For larger values
of $\zeta _{0}$ parameter $k_{1}$ becomes larger than $k_{2}$. This signals
that the blobs on the right hand side of the interface collapses to a size
smaller than the blobs on the left hand side, which is reasonable due to the
fact that the effect of self-interactions is amplified by the number of
monomers in the blob. However, this region is beyond the physical
realization of the present model, since we do not take into account the
repulsive third virial coefficient, which sets a minimal length for
collapsed blobs.

At some value of $\chi $ that corresponds to the temperature at which the
microphase separation occurs the parameter $k_{1}$ is always positive, so
that there is no delocalization transition (see curve 2 in Fig.2). Notice
that assuming the existence of a maximal value for the asymmetry parameter, $%
\zeta _{0}^{\max }$, the reentrant localization transition is limited to $%
\zeta _{0}<\zeta _{0}^{\max }$ , which imposes an inequality on $\chi $ and
thus on temperature.

Notice that the parameter $k_{1}$ becomes infinite at some value of $\chi $.
In the symmetric case ($\zeta _{0}=0$) we have found that $k_{1}$, $k_{2}$
become infinite at the critical value $\chi _{c}=9/(2\Delta )$ (Fig.3). The
latter means that the heteropolymer collapses onto the interface. Taking
into account the ternary interactions will prevent this.

\section{Conclusion}

We have considered the adsorption of a random heteropolymer onto an
interface within the model by Garel et al. \cite{gareletal89} by taking into
account the self-interactions between the monomers. The use of a
preaveraging procedure within the replica method permits to map the present
problem to a localization problem associated with a quantum mechanical
Hamiltonian. To find the binding state of the latter we use the variational
principle based on the 2-nd Legendre transform. We have found that
self-interactions favor the localization of the random heteropolymer driving
the delocalization transition to larger values of asymmetry. At the
appropriate strength of self-interactions we found a reentrant localization
transition at sufficiently high values of the asymmetry parameter $\zeta
_{0} $. Although we expect that the part of predictions are physically
irrelevant due to not taking into account the repulsive ternary
interactions, which will effectuate a minimal collapse length, the
qualitative tendencies of self-interactions will remain valid. The method we
use can be extended to go beyond the approximations used in the present
article. The preaveraging approximation means in fact a restriction of the
extremum functional associated with the second Legendre transform to
one-loop level, and can be overcomed. It is also possible to include the
longitudinal behavior of the heteropolymer into the extremum procedure.

\acknowledgments BVA thanks Deutscher Akademischer Austauschdienst (DAAD)
for fellowship. A support from the Deutsche Forschungsgemeinschaft (SFB 418)
is gratefully acknowledged.

\newpage Figure captions

Fig.1 Dependence of \ $k_{1}$ (solid line) and $k_{2}$ (dashed line) as a
functions of the asymmetry parameter $\zeta _{0}$ at $\chi _{0}=0$, $\Delta
_{0}=1$, $\sigma =1$, $T=1$ \cite{stepanowetal98}.

Fig.2 Dependence of \ $k_{1}$ (solid line) and $k_{2}$ (dashed line) as a
functions of the asymmetry parameter $\zeta _{0}$ at $\chi _{0}=1$, $\Delta
_{0}=1$, $\sigma =1$ (curves 1, 2 correspond to $T=0.5;1$ respectively).
Here LS, DS correspond to localized and delocalized states.

Fig.3 Dependence of \ $k_{1}$ (solid line) and $k_{2}$ (dashed line) as a
functions of the parameter $\chi $ at $\zeta _{0}=0$, $\Delta _{0}=1$, $%
\sigma =1$. $\chi _{c}$ is equal to $9/(2\Delta )$.

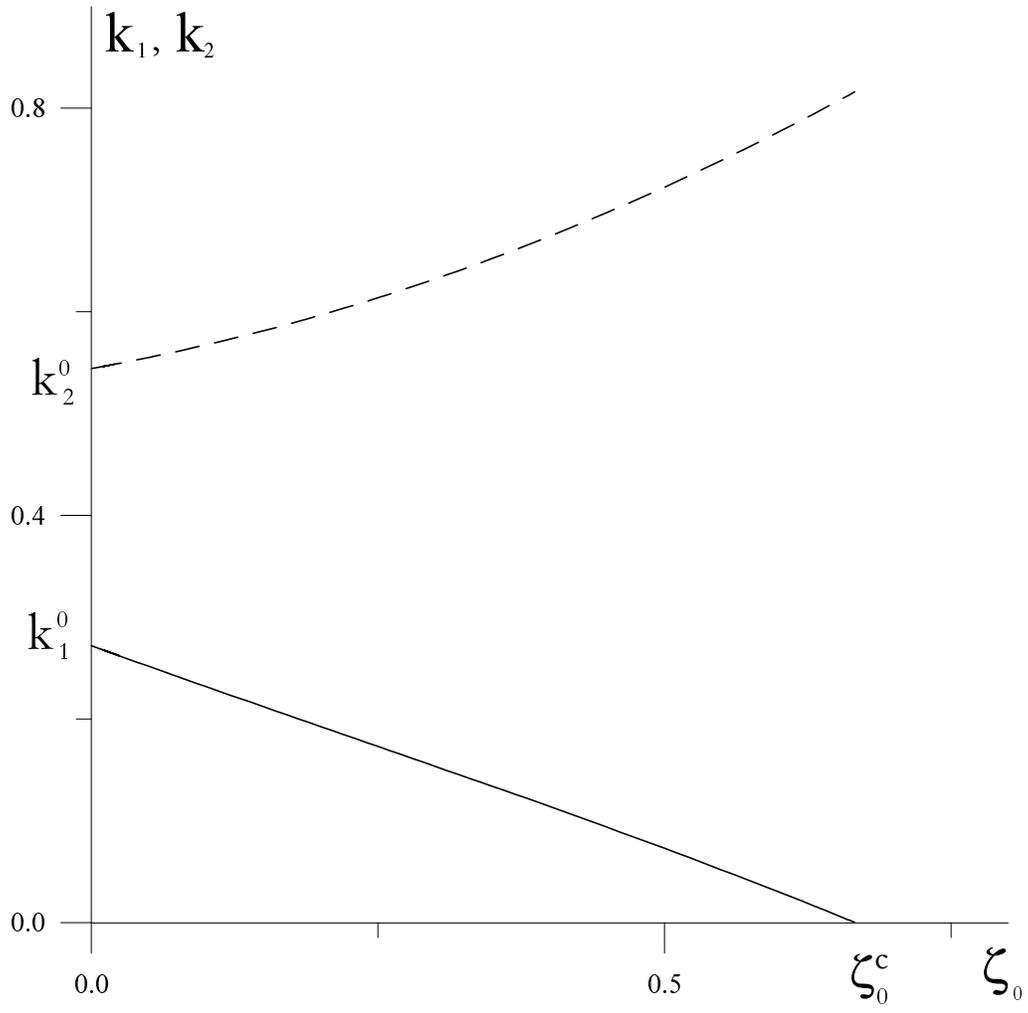
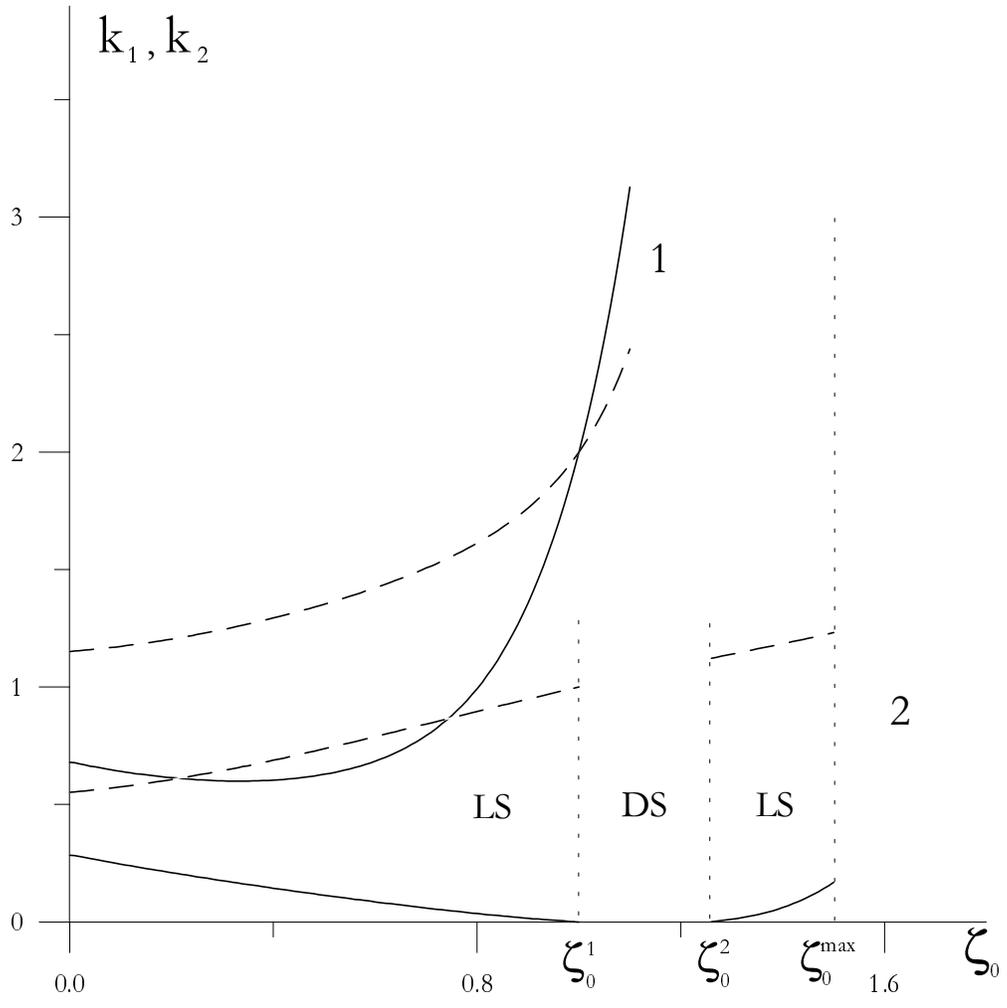
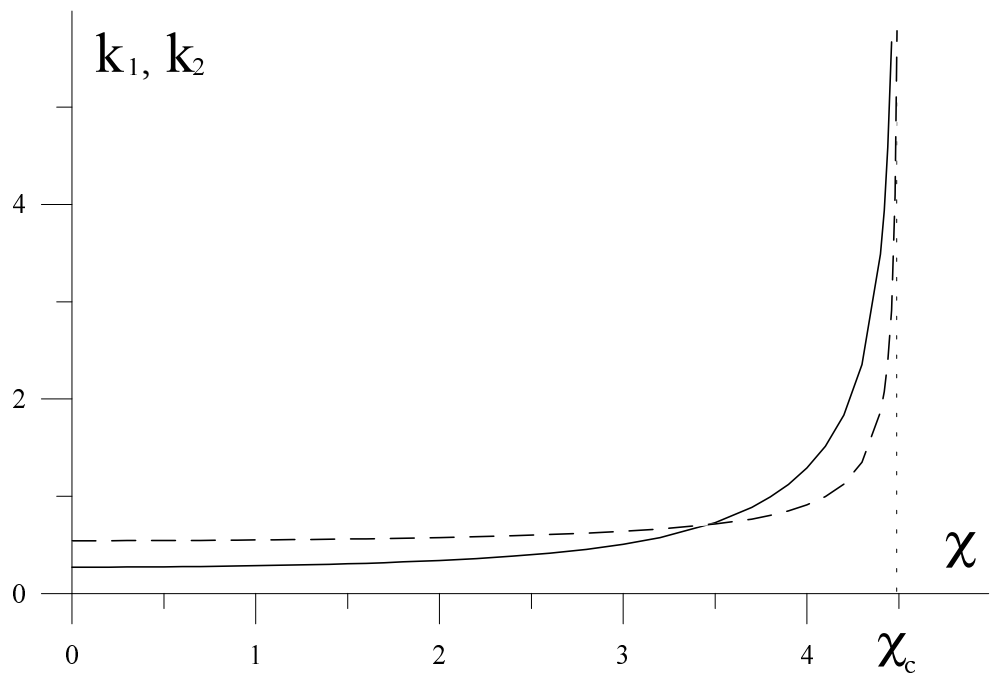
\end{document}